\documentclass[manuscript]{aastex}

\usepackage{graphicx}
\usepackage[space]{grffile}
\usepackage{latexsym}
\usepackage{amsfonts,amsmath,amssymb}
\usepackage{url}
\usepackage[utf8]{inputenc}
\usepackage{fancyref}
\usepackage[colorlinks=true,linkcolor=blue,citecolor=blue]{hyperref}
\usepackage{textcomp}
\usepackage{longtable}
\usepackage{multirow,booktabs}
\usepackage{slashbox}

\usepackage{natbib}

\def\dbd{death-by-dynamics} 
\def\Dbd{Death-by-dynamics}

\def\ptd{planetoid}

\def\sn1a{Type Ia supernovae}

\shorttitle{Death by Dynamics}  

\shortauthors{Rosanne Di\thinspace Stefano}

\begin{document}


\title{Death by Dynamics: Planetoid-Induced Explosions on White Dwarfs}  


\author{
Rosanne Di Stefano\affil{Harvard-Smithsonian Center for Astrophysics}
Robert Fisher\affil{University of Massachusetts Dartmouth} 
James Guillochon\affil{Harvard University} 
James F. Steiner\affil{Harvard-Smithsonian Center for Astrophysics}} 

\begin{abstract}
At intervals as short as
ten thousand years, each white dwarf (WD)  
passes within a solar radius of a \ptd , i.e., a comet, asteroid, or planet.
Gravitational tidal forces tear the \ptd\ apart; its
metal-rich debris falls onto the WD,  
enriching the atmosphere.
A third of WDs exhibit atmospheric ``pollution''.
For roughly every hundred \ptd\ disruptions, a \ptd\ collides 
with a WD. We simulate a small number
of collisions, in which ``death-by-dynamics'' refers to the fate
of the \ptd.  We also
compute the energies and likely durations of a broad sample of 
collision events,
and identify detection strategies 
at optical and X-ray wavelengths.    
Collisions with the most massive \ptd s   
can be detected in external galaxies. Some may trigger
nuclear burning. If one in  $\sim 10^7-10^8$ of WD-\ptd\ collisions creates the
conditions needed for a
Type Ia supernova (SN~Ia), 
``death-by-dynamics'' would also refer to the
fate of the WD, and 
 could provide a novel channel
for the production
of
 SN~Ia. We consider the circumstances
under which the rate of SNe~Ia can be increased by
interactions with \ptd s.
\end{abstract}




\section{Planetoids, Collisions with White Dwarfs, and the Generation
of Type Ia Supernovae}

\subsection{White Dwarfs and Tidal Disruptions}

Planets around white dwarfs (WDs) have yet to be discovered, but several
lines of reasoning suggest that WDs commonly host 
planetary systems (\citealt{RD.WDplanets}; \citealt{Parriott_Alcock_1998}). 
The strongest evidence is that 
more than $1/3$ of WDs  exhibit substantial
atmospheric ``pollution''. The mechanism 
responsible appears to be the tidal disruption (TD) 
of \ptd s (\citealt{Jura_2008}; \citealt{Zuckerman_2010}; \citealt{Farihi_Gansicke_Koester_2013b}). 
 
Because 
the metal settling time can be  
$\sim 10^5$ years 
(\citealt{Koester_2009}), contributions
from planetoids can be assessed only in aggregate.  
The average mass 
influx is estimated to be $\dot{M} = (10^{15}-10^{17})\,{\rm g/yr}$ 
\citep{Zuckerman_2010,Barber_2012}. Some WDs may therefore 
accrete up to an Earth mass of metal-rich material during a 
Hubble time.  

The mass is deposited by sequences of incident
planetoids: comets, asteroids, and planets. 
To relate $\dot M$ 
 to the influx of \ptd s, we 
use a power-law mass distribution, 
$\frac{dN(M)}{dM} \propto M^{-\beta}$, 
with $\beta =1.75$. 
Here, $dN(M)$
is the number of objects with mass between $M$ and $M+dM$.
Thus, the measured rate of pollution can be 
provided by the disruption of
 one $10^{12}$~g
TD per year, one $10^{15}$~g
 TD per century, or one $10^{21}$~g
 TD per Myr. The mass distribution
predicts a mix of masses and time intervals between events. 

\subsection{Collisions}

If \ptd s come close enough to WDs 
for TDs, then 
some must  collide with WDs. 
Because the cross section for close approaches is 
proportional
to the distance of closest approach,
the collision rate is $\sim 1\%$ the rate of
TDs, scaling roughly as the ratio $R_{WD}/R_\odot,$ 
where  $R_{WD}$ is the radius of the WD, and $R_\odot$ 
is very close to the tidal radius at which disruption 
occurs (e.g., \citealt{Pineault_Landry_1994};
\citealt{Schlichting_2013,Fuentes_2010}).
The word ``death'' 
in our title can therefore refer to the certain 
destruction of a planetoid
when it collides with a WD. 
In \S 2 we show the results of a small number of 
simulated collisions and then
compute the detectable effects of a larger number across the
mass range spanning from comets to planets.

\subsection{Type Ia Supernovae}

In extreme cases, a sequence of TDs and collisions  
may spark nuclear burning, potentially providing a
novel pathway to SNe~Ia. 
If interactions with \ptd s can help propel WDs to SN~Ia explosions,
the word ``death'' in our title may also refer to the end of the WD. 
The addition of a new evoluionary pathway may be a positive step 
toward developing a better understanding of how SNe~Ia are generated. 
Observations tell us that of every fifty WDs, one  
ends as a Type~Ia supernova (SN~Ia).  
Yet calculations based on the standard models have difficulty
producing rates this high \citep{Nelemans_2013}.

\section{Collisions between WDs and Planetoids}

\subsection{Energy of WD-Planetoid Collisions}

An incoming planetoid is subject to 
strong tidal forces once it crosses its tidal radius 
$r_{\rm t} = R_{\rm WD} (\bar{\rho}_{\rm WD}/\bar{\rho}_{\rm pl})^{1/3}$, 
where $\bar{\rho}_{\rm WD}$, $\bar{\rho}_{\rm pl}$ are the average densities 
of the WD and planetoid, respectively. At this moment, the planetoid disintegrates (likely into chunks) and has a spread in binding energies $\Delta E = G M_{\rm WD} R_{\rm WD} / r_{\rm t}^{2}$ that translates to a spread in times of impact $\Delta t \simeq (R_{\rm pl}/r_{\rm t}) t_{\rm ff} $, where $t_{\rm ff}$ is the amount of time elapsed between when the planetoid crosses $r_{\rm t}$ ($\simeq\,0.5\,R_{\odot}$ for $M_{\rm WD}\,=\,M_{\odot}$) and when it impacts the WD's surface, $t_{\rm ff} = \pi \sqrt{r_{\rm t}^{3}/8 G M_{\rm WD}} \sim 10~{\rm min}$. For a 4~km planetoid with $\bar{\rho}_{\rm pl} = 10~{\rm g}~{\rm cm}^{3}$ this spread would only be 7~ms, corresponding to an impactor whose length has increased by a factor of $\sim 5$.

\begin{figure*}
\centering\includegraphics[width=\linewidth,clip=true]{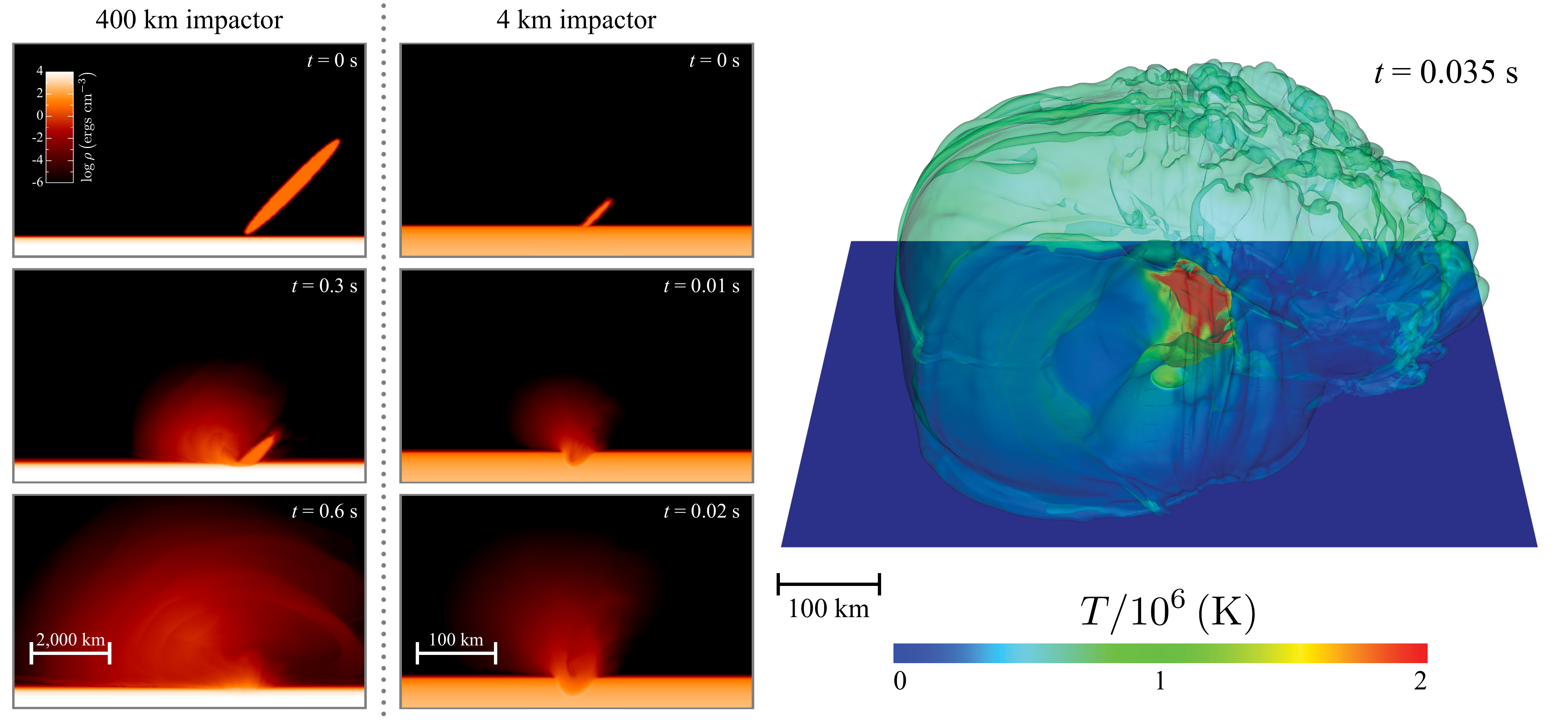}
\caption{Results from two hydrodynamical simulations of the impact of a tidally stretched bolide with a WD atmosphere. The left three panels show 2D slices of $\log \rho$ as a function of time $t$ through the 
plane of symmetry for an impactor initially 400 km in radius, where the impactor travels right to left at a 45$^\circ$ angle to the WD's surface. The middle three panels show the same for a 4 km impactor. The right panel shows a 3D snapshot from the 4 km impactor simulation, with the contour corresponding to the region that contains at least 10\% of the impactor's composition by mass, color-coded by temperature.}
\label{fig:hydro}
\end{figure*}

This modest increase in length means that the impactor strikes the surface of the WD in a small area, comparable to its own size, rather than being spread over the WD's surface \`a la the Shoemaker-Levy comet which impacted Jupiter on its second periapse \citep{Asphaug:1996a}. We simulate a two examples of planetoid impacts using the hydrodynamical code {\tt FLASH} \citep{Fryxell:2000a}, the results of which are presented in Figure \ref{fig:hydro}. We find that the planetoid penetrates to a depth where the density of the WD's atmosphere is a factor of $\sim 10$ greater than $\bar{\rho}_{\rm pl}$, with the surface and the impactor being heated to a few million degrees. After impact, the ejecta quickly expands and cools, initially radiating in the X-rays for $\sim$
 tens of milliseconds for the $4$~km impactor, 
longer for more
massive impactors, but releasing the majority of its energy in the optical over longer timescales.

\subsection{Event Detection}\label{subsec:detect}


State-of-the art optical surveys  -- e.g., Pan-STARRS \citep{Kaiser_2002}, Palomar Transient Factory (PTF; \citealt{Rau_2009}), SkyMapper \citep{Keller_2007}, GAIA \citep{Perryman_2001}, Palomar-Quest \citep{Djorgovski_2008}, Catalina Real-Time Transient Survey \citep{Drake_2009} -- can reach a 
field depth up to roughly $\sim$21 mag per observation 
(in one-to-several minutes exposure time;  \citealt{Rau_2009}, and references therein).  
We adopt the combined characteristics of Pan-STARRS and PTF as our benchmark.
Current surveys yield sky coverage up to several thousands of square 
degrees each 
night.  To classify a collision event as 
robustly detectable by present surveys, we adopt a strong 
threshold of 20~mag, in order that non-detections in data 
bracketing the transient event  can be  
discriminating.  To estimate an event's detectability, 
we adopt a time of $100$~s for a ground-based exposure 
 and neglect interstellar extinction.  We assume 
$100\%$ efficiency in the conversion of gravitational 
potential energy into optical radiation 
(for simplicity, we neglect any bolometric correction).  
The emission timescale is calculated following Arnett's 
law \citep{Arnett_1980} for an expanding gas.  
Recombination is the dominant source of emission, and 
continues until the gas has fully recombined, a timeframe which scales linearly
with the 
size of the impactor, with a rule of thumb $\sim 1$~s/km.  
The {\em effective} optical luminosity $\tilde{L}_{\rm opt}$ 
is the luminosity an observer would infer from a 100s 
exposure around the peak. 
Note that when the emission timescale becomes very short, 
the effective luminosity will be lower than the short-lived peak luminosity.

While optical surveys cover regions of the sky 
at a specified cadence, X-ray monitors 
(e.g., Swift BAT, \citealt{Barthelmy_2005}; 
RXTE ASM, \citealt{Levine_1996}; MAXI GSC, \citealt{Mihara_2011}; 
Integral IBIS, \citealt{Ubertini_2003}), often cover nearly the 
full sky continuously, producing $\sim$daily 
maps of the brightest X-ray sources.   One of the most-sensitive 
of these monitors, the MAXI GSC, can detect source 
fluences of $\sim50$ Crab~seconds per orbit 
(i.e., $\approx 10^{-6}$~erg~cm$^{-2}$, corresponding a persistent X-ray flux $\lesssim 50$ mCrab $\approx 10^{-9}$~erg~s$^{-1}$~cm$^{-2}$).  
MAXI observes very nearly the full sky each $\sim 90$~minute orbit.   
We adopt this fluence as a detection threshold.  
(Because the prompt emission is short compared to 
the exposure timescale, fluence and not flux is the critical quantity.)  
In calculating the X-ray emission, we arbitrarily assume 
that a fraction $f_X = 10\%$ of the impact 
energy $E$ is released in the form of prompt (nonthermal) X-rays.  

By contrast, Chandra / XMM-Newton imaging is 
sensitive to {\em much} fainter X-ray sources.  The tradeoff, of 
course, is that these instruments have very narrow fields-of-view, 
and a correspondingly low duty cycle for observing 
any particular region of sky.  We neglect photon pileup effects, 
and adopt a detection threshold of 25 X-ray photons (at a 
characteristic energy of 1~keV, for an instrument with 
effective area $\sim300$~cm$^2$).

In Table~1, we compute rough estimates for the rate, brightness, 
and detectability of impact events across a spectrum of masses.  The impactor masses
we consider range from $10^{14}$~g,
based on a lower optical threshold of $\tilde{L}_{\rm opt} = 10^{29}$~erg/s, 
to Jupiter's mass, $10^{30}$~g.  
In each case, we give the maximum distance for which the event 
could be detected in the optical and X-ray detectors discussed above. 

\begin{deluxetable}{lccccccl}
    \rotate
  \tabletypesize{\scriptsize}
  \tablecolumns{8}
  \tablewidth{0pc}
  \tablecaption{Event Rate and Detectability}
 \tablehead{\colhead{Mass} & \colhead{Rate} & \colhead{Peak $\tilde{L}_{\rm opt}$} &  \colhead{$f_X E /1 keV$}     &  \colhead{$d_{\rm opt}$} & \colhead{$d_{X,{\rm imag}}$} & \colhead{$d_{X,{\rm monitor}}$} & \colhead{Best Strategy}  \\ 
~~~~~~~~~~~~(g)                &  (yr$^{-1}$ Galaxy$^{-1}$) & (erg/s)   &     photons                &    (pc)       &  (pc)  & (pc) & }
\startdata

$10^{14}$  (103P/Hartley 2)        &  $8\times10^{6}$     &  $10^{29}$      &   $10^{39}$       &  150       &    8                          &  0.08   &  Ground-based optical surveys.\\ 
$10^{17}$  (SL9)     &  $4\times10^{4}$   &  $10^{32}$      &   $10^{42}$       &  5$\times10^3$      &    250                       &   2.5   &  Ground-based optical surveys. \\
$10^{19}$  (Eros)    &  $10^{3}$               &  $10^{34}$      &   $10^{44}$       &  50$\times10^3$      &    2.5$\times10^3$                      &  25   &  Ground-based optical surveys.\\
$10^{22}$  (Iris)       &  $8$                       &  $10^{37}$      &   $10^{47}$       &   1500$\times10^3$     &    80$\times10^3$           &   800   & X-ray all-sky monitors.  \\
$10^{24}$  (Ceres)   &  $0.25$            &  $10^{39}$      &   $10^{49}$       &   15$\times10^6$     &    800$\times10^3$         &    8$\times10^3$ &  X-ray all-sky monitors.   \\
$10^{26}$  (Mercury) &  $0.008$            &  $10^{41}$      &   $10^{51}$       &  150$\times10^6$      &    8$\times10^6$         &    80$\times10^3$ & X-ray imaging of galaxy clusters. \\
$10^{28}$  (Earth)   &  $2.5\times10^{-4}$ &  $3\times10^{42}$     &   $10^{53}$       &  800$\times10^6$      &    80$\times10^6$         &    800$\times10^3$ &  X-ray imaging of galaxy clusters. \\
$10^{30}$  (Jupiter) &  $8\times10^{-6}$   &  $2\times10^{44}$     &   $10^{55}$       &  7$\times10^9$      &   800$\times10^6$         &    8$\times10^6$ &   X-ray imaging / all-sky monitors. 
\enddata
\tablecomments{Estimates are subject to an approximate $\sim1$ order of magnitude uncertainty.}
\vspace{.25 cm} 
\label{tab:results}
\end{deluxetable}

Asteroids of mass $\lesssim 10^{20}$g produce the least energetic 
events in Table~1.  These are most readily detected in the optical, 
just within our Galaxy (or even just the Solar neighborhood).  
Although event rates are high, so that hundreds to thousands of 
events are {\em detectably} bright, these transient 
impacts are so brief that even high-cadence ground-based surveys 
designed to hunt transients are only likely to find 0.1--1 per 
decade (being most sensitive to impactors with $M\sim10^{19}$g).  

Meanwhile, energetic events involving sub-Earth-mass planets 
($M \sim 10^{27}$g) occur with sufficient frequency that they are 
potentially detectable with X-ray imaging of galaxy clusters.  In 
particular, several Ms aggregate imaging is available for crowded fields 
of nearby rich clusters ($d \lesssim 100$~Mpc, e.g., Virgo, 
Norma, and Coma).   Our model predicts $\lesssim 0.1$ events in 
the aggregate data set.  
However, we speculate that the vigorous dynamical activity in clusters may enhance these rates.

Intermediate impactor masses just below the scale of Ceres ($M \lesssim 10^{24}~g$), 
are sufficiently faint that they are all but undetectable outside the 
Local Group.  At the same time, the collision rates are so low that pointed 
ground or space-based instruments have negligible chance at serendipitous discovery. 
In this regime, the X-ray monitors shine.  Nearly 1 event
per year is detectably bright by an X-ray monitor; 
however, factoring in the sky coverage, 
only $0.01-0.1$/yr detections are expected.

Next decade, LSST, with its tenfold improvement in sensitivity 
is well-poised to make dozens of detections
of events by 
both the lowest-mass and highest-mass mass impactors.  
In particular, we predict that tens of Galactic events 
involving impactors with $M = (10^{14}-10^{18})$~g will be 
detected in the first year. 
The dominant sensitivity is for masses $M\sim10^{16}$~g, 
which roughly corresponds to the lowest-mass 
impactor for which the detectable volume is 
matched to the scale of the Galaxy.   
We predict tens of 
planet impact detections each year: we expect an annual 
rate of roughly one Mercury-scale detection, and nearly 
ten Jupiter-mass event detections. 
In the latter instance, the   
signal is dominated by high redshift events ($z \approx 1-3$). 

In short, we find that with current instruments, WD-planetesimal  
impacts are at the cusp  of detectability. It is unlikely that any 
event has heretofore been identified or recorded. At the same time, 
we note that with a generational improvement in sensitivity for 
X-ray coded-mask instruments, or in the forthcoming LSST at optical
wavelengths,  we will  
be well-poised to discover such impact events. Notably, a  modest 
improvement in  X-ray monitor sensitivity would critically extend 
the detectability of Jupiter-WD collisions out to distances of 
$\gtrsim 100$~Mpc, the proximity of nearby galaxy clusters.  
Detection rates for these most energetic events would then become   
appreciable. Even as we wait for a statistical sample of these 
events to be established using the rich capabilities of LSST,  
in this decade  monitoring programs by PTF, Pan-STARRS,  and 
other campaigns such as OGLE  will be increasingly capable of 
discovering a nearby \ptd -WD impact.

\section{The Progenitors of Type Ia Supernovae}

\subsection{Possibilities and Caveats} 

WD-\ptd\ collisions are frequent enough that if only one in $\sim 10^7-10^8$
produces an SN~Ia, \dbd\ models will
contribute significantly to the total rate. To trigger an 
SN~Ia,
a collision must dynamically ignite matter on a WD's surface.
This requires
compressing and heating 
material above a critical temperature
where the burning time becomes comparable to the dynamical time.
For pure hydrogen, the critical mass to auto-ignite a thin atmosphere
is small, and such WDs have very little in their atmospheres
(no more than $10^{-4}~M_{\odot}$). For pure helium, the maximum envelope mass
can be tenths of a solar mass for low-mass WDs
($\sim 0.5 M_{\odot}$), and decreases with increasing accretor mass \citep {bildstenetal07}.

If, however, a part of the envelope is compressed
and/or heated above
its steady-state condition, the fluid can spawn
a self-supporting nuclear detonation that propagates throughout the
envelope. Initiating such a detonation involves heating
a mass far smaller than the total envelope mass. For pure
helium, this mass is on the order of Earth's mass. Hence, we would
expect that, at minimum, the total impacting mass required to initiate
a detonation in the helium layer should be equal to this
value. Furthermore, the conditions for detonation are usually
realized at the interface between the outer envelope and the inner
core. In cases where the outer envelope mass is large,
an impactor would need to penetrate through a mass greater than
its own to
reach the depth of this interface. Secondly, planetary/planetesimal
impactors are low-density, an Earth-mass body has roughly the
same size as the WD, and its kinetic energy
is spread over a large shell (meanwhile, the critical length scales
are derived presuming all the mass lies within a small, spherical volume).

We envision two circumstances which could change this picture. Firstly,
the critical length scales have only been calculated for
a few composition combinations, namely carbon-oxygen \citep {seitenzahletal09},
and pure helium \citep {holcombetal13}. However, the inclusion of even a small admixture
of heavy nuclei into a helium layer will greatly reduce the burning timescale, since $\alpha$ captures
onto these nuclei are far more rapid than the relatively slow 3-$\alpha$
reaction \citep {shenmoore14}. Consequently, while a single impact
may not detonate a pure helium envelope, repeated
bombardment may contaminate the envelope to such
a degree that $\alpha$ captures and other nuclear pathways become viable.
Secondly, because the impact is fundamentally a dynamical process,
the geometry of the impact may lead to a shock-focused detonation in a layer
stable to a single hot spot generated by the collision -- see e.g., \citet {finketal10}.

\subsection{Death by Dynamics in Tandem with SD and DD models}

The two leading models for SN~Ia progenitors invoke interacting binaries
with epochs of mass transfer and common envelope evolution. These epochs
last long enough 
to permit sequences of \ptd\ interactions.
In single degenerate (SD) models,
a non-degenerate companion donates matter to the WD.
In double degenerate (DD) models
two WDs merge, leading to a detonation.

Planetoid interactions that enrich the WD may alter SN~Ia rates
in three ways. 
Firstly, 
if enrichment extends (narrows) the range
of infall rates compatible with steady nuclear 
burning \citep {piersantietal14}, then both DD and SD rates would
be affected. Secondly, enhanced metallicity increases winds, possibly
stabilizing  
mass transfer for
systems verging on
dynamical instability. Finally, enrichment may make it easier or
harder to spark explosions during the accretion/merger processes 
associated with explosions in SD/DD models.  
Furthermore, a subset of \ptd\ interactions
with the donor, disk,
or common envelope could alter the binary's further evolution. 

\subsection{Predictions of \Dbd\ Models}

Death-by-dynamics models predict diversity among SNe~Ia.
The WD mass 
does not have a fixed value at explosion.
Differences in \ptd\ compositions produce    
chemical diversity.
These models are novel in positing that a
 WD can become an SN~Ia without a close companion.  
There may, however, be other stars near the WD: 
distant bound companions, passing stars which
have triggered \ptd\ interactions, or even close interacting companions.   
Thus, the properties of any stars in the field close to the SN
may exhibit a variety of luminosities and spectral types.
The circumstellar region {may be}  
enriched 
with planetary 
debris that interacts with the supernova. 

Early-time explosions are predicted by \dbd\ models, 
because stellar evolution destabilizes \ptd\ orbits. 
As well, planetoids 
continue to be placed on eccentric orbits
over a Hubble time. 
The combination of strong early-time signals and
continuation through late times is consistent with
data on SNe~Ia (e.g., \citet{Sullivan.2006}). 
Calculations are needed to
check whether the distribution of \dbd\ delay times is
consistent with observations.

The discovery of WD-\ptd\ collision events
will identify 
environments that produce SNe~Ia through \dbd\ channels.
Conversely, regions
producing SNe~Ia must be rich in
WD-\ptd\ interactions.


\section{Conclusions} 
 
WD-\ptd\ collisions produce bright, detectable events.
Observing these events 
will measure peak luminosities and time evolution
in broad energy bands, and will quantify  
environmental effects.
This will provide clues to the mass distributions of
both  planetoids 
and WDs, and eventually 
chemical information.
Such studies are possible because large numbers of \ptd s
orbit WDs and large numbers of WDs inhabit galaxies. 
Even unusual interactions and sequences of interactions
are common in galaxies and
galaxy clusters.

A \ptd\ interaction or sequence of them
may spark SNe~Ia. The exploding WDs may be isolated or
may be in interacting binaries.
While it will be challenging to test \dbd\ SNe~Ia
progenitor models,
we note that this has also been true for SD and DD models. 
Fortunately, 
if \dbd\ models do produce SNe~Ia, they make a host of predictions, 
some of which 
can be tested through the study 
of bright events caused by WD-\ptd\ 
collisions. 
In summary, \dbd\ models make testable predictions and suggest intriguing directions
for future investigations. 

\smallskip
\noindent{\bf Acknowledgements:} 
RD thanks Stuart Sim and R\"{u}diger Pakmor for enlightening discussions. 
This work was supported 
by NSF AST-1211843,
AST-0708924, AST-0908878 and
NASA NNX12AE39G, AR-13243.01-A.
JFS was supported by the NASA Hubble Fellowship HST-HF-51315.01. 
JG was supported by the NASA  
Einstein Fellowship PF3-140108.

\bibliographystyle{apj}

\bibliography{wd_planetoids}

\end{document}